\documentstyle[art10]{article}
\newcommand{\be}{\begin{equation}}
\newcommand{\ee}{\end{equation}}

\begin{document}
\newcommand{\plt}{Phys.Lett.}
\begin{center}
{\Large{\bf CONSTITUENT GLUONS FROM QCD}}\\[0.2cm]
Yu.S.KALASHNIKOVA\\
{\it Institute of Theoretical and Experimental Physics
, \\117259 Moscow, Russia}%
\\[0.5cm]

\end{center}
\footnotesize
The notion of constituent gluon is introduced as a gluon propagating
in the vacuum background field.
 The Hamiltonian approach for a system containing such gluon and a
$q\bar{q}g$ pair is formulated, and the masses of lowest $q\bar{q}g$
hybrids are estimated.\\[0.5cm]
 \normalsize

In the QCD--motivated constituent models not only quarks but also
gluons should be confined, so that the states containing constituent
glue should exist. The data on mesonic spectra and decays in the mass
region 1.5-2.0 GeV indicate at the existence of "extra" states, or
the states with properties uncompatible with properties of ordinary
$q\bar q$ mesons. It seems reasonable to study various theoretical
models to find out if the gluonic hadrons are admissable in their
framework, and if the predictions have something to do with the
experimental data.

 Here the studies of hybrid mesonic excitations are
 presented in the framework of Vacuum Background
 Correlator method [1]. The constituent gluon is
 introduced starting from the perturbation theory
 expansion in nonperturbative QCD vacuum [2]. The gluonic field
 $A_{\mu} $ is split into the background part $B_{\mu}$ and the
 perturbation $a_{\mu}$ over background, and,
 for example, one--gluon hybrid may be
 represented as
 \be
 \Psi(x_q,x_{\bar q},x_g)=
 \psi(x_{\bar q})\Phi(x_{\bar q}x_g)a(x_g)\Phi(x_gx_{\bar
 q})\psi(x_q)
 \ee
 where parallel transporter $\Phi$ contains only background field.

 The Green function of a $q\bar q g$ state can be written using the
 Feynman--Schwinger representation as a path integral over paths of
 quarks and gluon: (the details may be found in [3]):
 \be
  G(x_qx_{\bar q}x_g; y_qy_{\bar q} y_g)=
  \ee
  $$
  \int^{\infty}_0ds
  \int^{\infty}_0d\bar s
  \int^{\infty}_0dS \int DzD\bar zDZ exp(-{\cal{K}}_q-{\cal{K}}_{\bar
  q}-{\cal{K}}_g)<{\cal{W}}>_B
  $$
  where
  $$
  {\cal{K}}_q=m^2_qs+\frac{1}{4}\int^s_0\dot z^2(\tau)d\tau,~~
  {\cal{K}}_{\bar q}=m^2_{\bar q}\bar s+\frac{1}{4}\int^{\bar
  s}_0\dot{\bar z}^2(\tau)d\tau,~~
  {\cal{K}}_g=\frac{1}{4}\int^{\xi}_0\dot{Z}^2(\tau)d\tau,
  $$
   and the  Wilson loop operation ${\cal{W}}$ in (2) may be written
   as
   \be
   {\cal{W}}=\frac{1}{2} W_1W_2-\frac{1}{2N_c}W, \ee where $W_1,
   W_2$ and $W$ are the Wilson loops (in the fundamental
   representation) along the closed contours  formed by the paths of
   quark and gluon, antiquark and gluon, and quark and antiquark
   correspondingly.

   To average the Wilson loop configuration (3) one may use the
   cluster expansion  method [4] with the result
   \be
   <{\cal{W}}>_B=\frac{N_c^2-1}{2} exp -\sigma(S_1+S_2))
   \ee
   for large contours, where $S_1$ and $S_2$ are the minimal areas
   inside the contours formed by paths of quark and
   gluon and antiquark and gluon.

   To formulate the Hamiltonian approach one should extract the
   effective Lagrangian from (2), and reduce the four--dimensional
   dynamics to the three--dimensional one. Assuming the straight--
   line approximation for the minimal surfaces, one arrives to the
   Hamiltonian for the $q\bar q g$ system, which for low orbital
   momenta takes the form
   \be
   H=\frac{m_q^2}{2\mu_1}+ \frac{m_{\bar q}^2}{2\mu_2}+
\frac{\mu_1+\mu_2+\mu_3}{2}+\frac{p^2}{2\mu_p}+\frac{Q^2}{2\mu_{Q}}+
   \sigma\rho_1+\sigma\rho_2,
   \ee
   where the Jacobi coordinates $\vec{r}$ and $\vec{\rho}$ and
   conjugated momenta $\vec{p}$ and $\vec Q$ are introduced,
   $$\mu_p=\frac{\mu_1\mu_2}{\mu_1+\mu_2},~~
   \mu_Q=\frac{\mu_3(\mu_1+\mu_2)}{\mu_1+\mu_2+\mu_3},
   $$
   and the quantities $\mu_i$ are the fields over which one  is to
   integrate in the path integral representation, or, equivalently,
   to take extremum in $\mu_i$ in the Hamiltonian. It  appears,
   however, that one may first find the eigenvalues of the
   Hamiltonian (5) assuming $\mu_i$ to be $c$-numbers, and then
   minimize the eigenvalues in $\mu_i$, considering, in such a way,
   $\mu_i$ as constituent masses  of quarks  and gluon. This
   procedure works with rather good accuracy for the lowest states.

   The physical states are defined to be transverse with respect to
   the gluon  momentum
, so
that the possible quantum numbers for the ground state hybrids are
   \be
   J^{PC}=0^{\mp +},1^{\mp +},2^{\mp +},1^{\mp -}.
   \ee
   The actual calculations of the mass spectra were carried on with
   the inclusion of Coulomb force  with $\alpha_s=0.3$ and the values
   of quark masses were chosen to be $m_q=0.1 GeV, m_s=0.25 GeV,
   m_c=1.5 GeV.$ The absolute mass scale was set by adding the
   constant term to the Hamiltonian, which was chosen to be twice as
   in the corresponding $q\bar q$ system. The results for the masses
   of ground state hybrids read:
   \be
   M(q\bar q g)=1.7 GeV,~~ M(s\bar s g) = 2.0 GeV,~~
   M(c\bar c g)=4.1 GeV.
   \ee

   The values (7) appear to be very close to the mass range where the
   hybrid candidates are believed to be placed [5], and
   to the values obtained in the flux tube model [6].
   The latter has a lot of common with the present
   approach: both models take into account the
   transverse motion of the string.  The main difference
   is in quantum numbers: in the present model a
   perturbative gluon which carries it's own quantum
   numbers is needed to make  a string vibrate.

   At present several hybrid candidates are under
   discussion [5], with hybrid assignement suggested not
   only because of the values of masses, but because of
   their decay properties: the decay of hybrid into two
   $S$--wave mesons is suppressed in the flux tube model. Just
   the  same signature exists for a hybrid with electric
   constituent gluon [7]. So more sophisticated selection rules
   involving spin content of the decay products [7] are needed
   to tell the flux tube from constituent hybrid.

\noindent {\bf Acknowledgments}\\[0.10cm]
This work is supported in part by the Grant NJ77100 from the
International science Foundation and Russian Government, and by the
Grant INTAS -93-0079.
\\[0.15cm]
\noindent {\bf References}\\[0.10cm]
1. H.G.Dosh and Yu.A.Simonov, Phys. Lett. B. B 205, 339 (1988)\\
{}~~~Yu.A.Simonov, Nucl. Phys. B 307, 512 (1988); B324, 67 (1989)
\\
2. L.F.Abbot, Nucl.Phys. B 185, 189 (1981)\\
{}~~~Yu.A.Simonov, HD-THEP-93-16
\\
3. Yu.S.Kalashnikova and Yu.B.Yufryakov, Preprint ITEP-35-95,
hep-ph/9506269,\\
{}~~~Phys.Lett.B (in press)
\\
4. A.Yu.Dubin, Yu.S.Kalashnikova, Preprint ITEP-40-95,
hep-ph/9406332,\\
{}~~~Yad.Fiz. (in press)
\\
5. F.Close, these proceedings.
\\
6. T.Barnes, F.E.Close, E.S.Swanson,  ORNL-CTP-95-02, RAL-94-106,
\\~~~hep-ph/9501405\\
{}~~~E.S.Swanson, these proceedings.
\\
7. A.Le Yaonanc, L.Oliver, O.Pene, J.C.Raynal, Z.Phys. C28, 309 (1985)
   \\
   ~~~Yu.S.Kalashnikova, Z.Phys. C62, 323 (1994)

 \end{document}